\documentclass[fleqn,12pt]{article}
\usepackage{graphics,epsfig,here}

\def\beq   {\begin{equation}}
\def\eeq   {\end{equation}}
\def\beqd  {\begin{displaymath}}
\def\eeqd  {\end{displaymath}}
\def\beqaa {\begin{eqnarray}}
\def\eeqaa {\end{eqnarray}}

\def\ti  {\tilde}

\def\sz{\ifmmode{\tilde{\chi}^0} \else{$\tilde{\chi}^0$} \fi}
\def\sw{\ifmmode{\tilde{\chi}} \else{$\tilde{\chi}$} \fi}

\newcommand{\be}[1]{\begin{equation} \label{(#1)}}
\newcommand{\ee}{\end{equation}}
\newcommand{\baq}[1]{\begin{eqnarray} \label{(#1)}}
\newcommand{\eaq}{\end{eqnarray}}
\newcommand{\rf}[1]{(\ref{(#1)})}
\newcommand{\ba}{\begin{array}}
\newcommand{\ea}{\end{array}}

\newcommand{\slashed}[1]{\not\!#1}

\begin{document}
\pagestyle{empty}

\vspace*{-1cm} 
\begin{flushright}
  UWThPh-2004-07 \\
  hep-ph/0403265
\end{flushright}

\vspace*{1.4cm}

\begin{center}

{\Large {\bf CP sensitive observables in chargino production
with transverse $e^{\pm}$ beam polarization
}}\\

\vspace{2cm}

{\large 
A.~Bartl, K.~Hohenwarter-Sodek, T.~Kernreiter, H.~Rud}

\vspace{2cm}

{\it Institut f\"ur Theoretische Physik, Universit\"at Wien, A-1090
Vienna, Austria}

\end{center}

\vspace{2cm}

\begin{abstract} 
We consider the process $e^+e^-\to\ti\chi^+_i\ti\chi^-_j$
at a linear collider with transverse $e^{\pm}$ beam polarization.
We investigate the influence of the CP phases
on azimuthal asymmetries in $e^+e^-\to\ti\chi^+_i\ti\chi^-_j$ 
with subsequent two-body decays $\ti\chi^-_j\to \ti\nu_{\ell}\ell^-$ 
and $\ti\chi^-_j\to W^-\ti\chi^0_1$.
We show that triple product correlations involving the
transverse $e^{\pm}$ beam polarization vanish if at least
one subsequent chargino decay is not observed. 
We derive this result within the Minimal Supersymmetric 
Standard Model (MSSM) with complex parameters, 
however, it holds also in the general
MSSM with SUSY flavour violation.

\end{abstract}

\newpage
\pagestyle{plain}


\section{Introduction}

Supersymmetry (SUSY) is at present the most studied
extension of the Standard Model (SM) \cite{Haber:1984rc}.
Some of the SUSY parameters may be complex and
are potential new sources of CP violation \cite{Dugan:1984qf,Masiero:xj}. 
In the chargino sector of the Minimal Supersymmetric Standard 
Model (MSSM) the higgsino mass parameter $\mu$ 
can be complex, while the $SU(2)$ gaugino mass parameter 
$M_2$ can be chosen real by redefining the fields.
In the neutralino sector of the MSSM also 
the gaugino mass parameter $M_1$ can be complex.
The precise determination of the underlying SUSY
parameters will be one of the main goals of a future
$e^+e^-$ linear collider (LC) with high luminosity  \cite{TDR}.
The phases $\varphi_{\mu},\varphi_{M_1}$ will give rise 
to CP-odd observables which may also be measured 
in future collider experiments.

The study of chargino production 
\be{eq:prodchar}
e^+e^-\to\ti\chi^+_i\ti\chi^-_j ~,\qquad i,j=1,2~,
\ee
will play an important role at the LC. This process has been studied quite often 
in the literature \cite{Bartl:fk,Choi:1998ei,Choi:2000ta}. 
In \cite{Choi:1998ei} a method has been developed to determine
the underlying parameters $M_2, |\mu|, \tan\beta,$ including
$\cos\varphi_{\mu}$, by measurements of the chargino
masses and cross sections. The formulae for the cross
section of \rf{eq:prodchar} including longitudinal
and transverse beam polarizations have also been given
and azimuthal asymmetries have been proposed 
in \cite{Choi:2000ta,Chiappetta:1985nb}.

In principle experiments with transverse $e^{\pm}$
beam polarization may offer the possibility of precision
studies of the effects of CP violation and new physics.
For example, it has been shown that in the reactions
$e^+e^-\to W^+ W^-$ \cite{Fleischer:1993ix}, 
$e^+e^-\to f ~\bar f$ \cite{Budny:tk} and 
$e^+e^-\to t ~\bar t$ \cite{Ananthanarayan:2003wi}
transverse $e^{\pm}$ beam polarization is indeed
very helpful to disentangle effects of new physics.
It is, therefore, tempting to study the potential of
transverse beam polarization for measuring
CP sensitive observables also in chargino production \rf{eq:prodchar}.
Triple products give rise to T-odd observables which may be useful
to measure the CP phases involved.
When only the cross sections of \rf{eq:prodchar} are
measured, summed over the polarizations of the produced
charginos, one may try the following triple products
involving the transverse beam polarization
\be{eq:triple1}
O^1_T =
({\bf p}_e\times{\bf p}_{\ti\chi})\cdot{\bf t^{\pm}}~,\qquad
O^2_T =({\bf t^+}\times{\bf t^-})\cdot{\bf p}_{\chi} ~,
\ee
where ${\bf t^-}$ (${\bf t^+}$) is the three-vector of the
transverse polarization of the $e^-$ ($e^+$), and
${\bf p}_e$ and ${\bf p}_{\ti\chi}$ are the momentum vectors
of $e^-$ (or $e^+$) and $\ti\chi^{\pm}_i$.
The leading contribution (at tree-level) to such a
term in the matrix element would be solely due to CP violation.
However, from the formulae given in \cite{Choi:2000ta}
it can be seen that terms involving the triple products $O^{1,2}_T$ vanish
if only chargino production cross sections are measured.
This follows also from the general analysis in 
\cite{Ananthanarayan:2003wi} and \cite{Dass:1975mj}.
As a next step one may try to use triple products
which involve also the subsequent decay of one of the 
two charginos.
For definiteness one may consider the two-body decays 
\be{eq:lepdec}
\ti\chi^-_j\to \ti\nu_{\ell}\ell^-
\ee
and
\be{eq:Wdec}
\ti\chi^-_j\to W^-\ti\chi^0_1~.
\ee
Then the momentum vector of $\ell=e,\mu,\tau$
or the $W$ boson (if $W$ decays hadronically) 
may be used to study the following triple products: 
\be{eq:triple2}
O^3_T =
({\bf p}_e\times{\bf p}_{\ell,W})\cdot{\bf t}^{\pm}~,\qquad
O^4_T =
({\bf t}^+\times{\bf t}^-)\cdot{\bf p}_{\ell,W}~.
\ee
Possible T-odd observables based on the triple products in
Eqs.~\rf{eq:triple1} and \rf{eq:triple2}
would be
\be{eq:Todd}
\langle O^i_T\rangle~,\quad \langle{\rm sgn}(O^i_T)\rangle~, \quad 
i=1,\dots,4~.
\ee
However, as we will show below also the T-odd observables \rf{eq:Todd} vanish in Born
approximation and neglecting terms proportional to the electron mass. 

In the present paper we examine again the triple product correlations
Eqs.~\rf{eq:triple1} and \rf{eq:triple2}. 
We give a further argument why they have to vanish. In order
to make use of the transverse beam polarization in chargino production
we define the azimuthal asymmetries for the cases:
($i$) Azimuthal distribution of $\ti\chi^-_j$ (when the 
direction of flight of the charginos can be reconstructed).
($ii$) Azimuthal distribution of the decay product in the decays
\rf{eq:lepdec} or \rf{eq:Wdec}.
As we will demonstrate, the azimuthal asymmetry, though CP even,
may serve as a good observable to study the effects of CP phases.

The paper is organized as follows:
In section \ref{sec:2} we present the formulae for 
the cross section of \rf{eq:prodchar} with transverse beam polarization and the
decays \rf{eq:lepdec} and \rf{eq:Wdec}.
In section \ref{sec:3} we argue why the T-odd observables
in \rf{eq:Todd} vanish if only one of the subsequent chargino decays 
is considered.
We define in Section \ref{sec:4} the azimuthal asymmetries 
and present our numerical results.
Section \ref{sec:5} contains our conclusions.

\section{Cross section\label{sec:2}}

\begin{figure}[t]
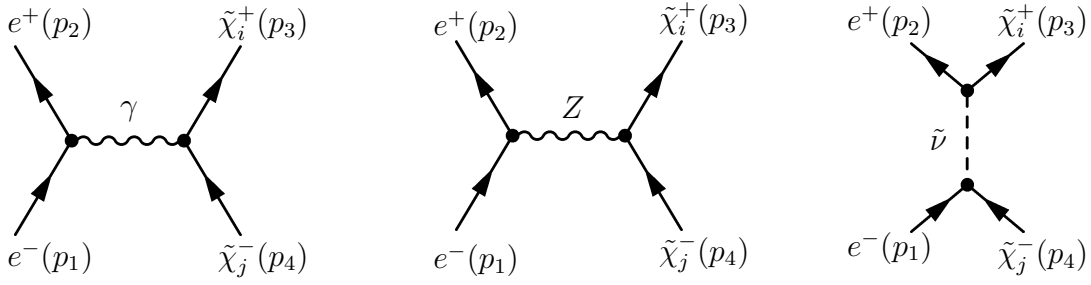

\hspace{1cm}
\begin{minipage}[t]{3.5cm}
\begin{center}
{\setlength{\unitlength}{1cm}
\begin{picture}(1.5,2.5)
\put(-2.,-1.1){\includegraphics{fig.1a}}
\put(-1.8,-1.6){$e^{-}(p_1)$}
\put(1,-1.6){$\tilde{\chi}^{-}_j(p_4)$}
\put(-1.8,1.5){$e^{+}(p_2)$}
\put(1,1.5){$\tilde{\chi}^{+}_i(p_3)$}
\put(-.3,.4){$\gamma$}
\end{picture}}
\end{center}
\end{minipage}
\hspace{2cm}
\vspace{.8cm}

\begin{minipage}[t]{3.5cm}
\begin{center}
{\setlength{\unitlength}{1cm}
\begin{picture}(-2.5,2.5)
\put(3.,2.3){\includegraphics{fig.1a}}
\put(3.,1.7){$e^{-}(p_1)$}
\put(6.,4.9){$\tilde{\chi}^{+}_i(p_3)$}
\put(3.,4.8){$e^{+}(p_2)$}
\put(6.,1.8){$\tilde{\chi}^{-}_j(p_4)$}
\put(4.7,3.7){$Z$}
\end{picture}}
\end{center}
\end{minipage}
\hspace{2cm}
\vspace{.8cm}

\begin{minipage}[t]{3.5cm}
\begin{center}
{\setlength{\unitlength}{1cm}
\begin{picture}(2.5,2.5)
\put(11.8,5.4){\includegraphics{fig.1b}}
\put(11,8.2){$e^{+}(p_2)$}
\put(11,5.2){$e^{-}(p_1)$}
\put(13,8.2){$\tilde{\chi}^{+}_i(p_3)$}
\put(13,5.1){$\tilde{\chi}^{-}_j(p_4)$}
\put(12.1,6.6){$\tilde{\nu}$}
 \end{picture}}
\end{center}
\end{minipage}
\vspace{-3.8cm}
\caption{\label{bild1} 
Feynman diagrams for $e^+e^-\rightarrow\tilde{\chi}^+_i\tilde{\chi}^-_j$ }
\end{figure}

The Feynman diagrams contributing to process \rf{eq:prodchar} 
are given in Fig.~\ref{bild1}.
The relevant parts of the interaction Lagrangian 
which contribute to the process 
$e^+e^-\rightarrow\tilde{\chi}^+_i\tilde{\chi}^-_j$ and the
subsequent two-body decays
$\ti\chi^-_j\to \ti\nu_{\ell}\ell^-$ and
$\ti\chi^-_j\to W^-\ti\chi^0_1~$
are given by \cite{Haber:1984rc}
\begin{equation}
\mathcal{L}_{Z e^- e^+}=-
\frac{g}{\cos\Theta_W}Z_{\mu}{\bar \psi}_e\gamma^{\mu}
(L_e P_L+R_e P_R) \psi_e~,
\end{equation}
\begin{equation}
\mathcal{L}_{Z\tilde{\chi}^+\tilde{\chi}^-}=
\frac{g}{\cos\Theta_W}Z_{\mu}\overline{\tilde\chi^+_i}\gamma^{\mu}
(O'^L_{ij}P_L+O'^R_{ij}P_R){\tilde{\chi}}^+_j~,
\end{equation}
\be{eq:lagsneuchar}
\mathcal{L}_{\ell\tilde{\nu}\tilde{\chi}^+}=
-gV^*_{j1}\overline{\tilde{\chi}^{+C}_j}P_L 
\ell \ti\nu^*+{\rm h.c.}~,
\ee
\be{eq:lagneuchar}
\mathcal{L}_{W^-\tilde{\chi}^+\tilde{\chi}^0}=
g W^-_{\mu} \overline{\tilde{\chi}^0_k}\gamma^{\mu}(O^L_{kj} P_L+O^R_{kj} P_R) 
\tilde{\chi}^+_j+{\rm h.c.}~,
\ee
with
\begin{equation}
O'^L_{ij}=-V_{i1}V^*_{j1}-\frac{1}{2}V_{i2}V^*_{j2}+
\delta_{ij}\sin^2\Theta_W~,
\end{equation}
\begin{equation}
O'^R_{ij}=-U^*_{i1}U_{j1}-\frac{1}{2}U^*_{i2}U_{j2}+
\delta_{ij}\sin^2\Theta_W~,
\end{equation}
and
\be{eq:OLOR}
O^L_{kj}=-\frac{1}{\sqrt{2}} N_{k4}V^*_{j2} + N_{k2}V^*_{j1}~,\qquad
O^R_{kj}=\frac{1}{\sqrt{2}} N^*_{k3}U_{j2} + N^*_{k2}U_{j1}~,
\ee
where $L_e=-1/2+\sin^2\Theta, R_e=\sin^2\Theta$,
$P_{L,R}=1/2(1\mp\gamma_5)$, 
$g$ is the weak coupling constant, $e=g \sin\Theta_W$ 
and $\Theta_W$ is the Weinberg angle. 
The unitary $2\times2$ mixing matrices $U$ and $V$ 
diagonalize the chargino mass matrix ${\mathcal M}_C$, 
$U^{\ast}{\mathcal M}_C V^{-1}=
{\rm diag}(m_{\chi_1},m_{\chi_2})$.
$N_{ij}$ is the complex unitary $4\times 4$ matrix which diagonalizes
the neutral gaugino-higgsino mass matrix $Y_{\alpha\beta}$, 
$N_{i \alpha}^*Y_{\alpha\beta}N_{k\beta}^{\ast}=
m_{\chi^0_i}\delta_{ik}$,
in the basis ($\tilde{B},
\tilde{W}^3, $ $ \tilde{H}^0_1, \tilde{H}^0_2$) \cite{Haber:1984rc}.

For the calculation of the amplitude squared of the 
process \rf{eq:prodchar} with subsquent decays \rf{eq:lepdec} 
and \rf{eq:Wdec}, 
we use the spin-density matrix formalism \cite{bouchiat,spinhaber}. 
The amplitude squared (without summing over the
polarization of the charginos) can be written in 
the following way
\baq{eq:spindensity}
\rho_P^{\lambda_i\lambda'_i\lambda_j\lambda'_j} &=&
\delta_{\lambda_i\lambda'_i}\delta_{\lambda_j\lambda'_j}P
+\delta_{\lambda_j\lambda'_j}\sum_a\sigma^a_{\lambda_i\lambda'_i}
\Sigma^a+\delta_{\lambda_i\lambda'_i}
\sum_b\sigma^b_{\lambda_j\lambda'_j}\Sigma^b
\nonumber\\[3mm]
 & &+\sum_{ab}\sigma^a_{\lambda_i\lambda'_i}
\sigma^b_{\lambda_j\lambda'_j}
\Sigma^{ab}~,
\eaq
where the coefficient $P$ represent the part of the
amplitude squared which is independent 
of the polarization of the $\tilde{\chi}^{\pm}$'s, and $\Sigma^a$
and $\Sigma^b$ contain the parts which  
depend on the polarization of $\tilde{\chi}_i^+$ and
$\tilde{\chi}_j^-$, respectively.
Finally $\Sigma^{ab}$ contains the part which depends
on the polarization of both $\tilde{\chi}^{\pm}$'s.
In Eq.~\rf{eq:spindensity} $\sigma^{a,b}$ ($a,b=1,2,3$) denote
the Pauli matrices and $\lambda_i,\lambda_i'$ ($\lambda_j,\lambda_j'$)
are the helicity indices of $\tilde{\chi}_i^+$ ($\tilde{\chi}_j^-$).

In the treatment of beam polarizations we use the general parametrization which,
in the limit of vanishing electron mass, $m_e\to 0$, is given by
\be{eq:proje}
\lim_{m_e\to 0}\frac{1}{2}({\bf 1}+\gamma_5\slashed s_{e^-})(\slashed p_1 + m_e)=
\frac{1}{2}({\bf 1}+P_L\gamma_5+\gamma_5 P_T \slashed t^-)\slashed p_1
\ee
and
\be{eq:projp}
\lim_{m_e\to 0}\frac{1}{2}({\bf 1}+\gamma_5\slashed s_{e^+})(\slashed p_2 - m_e)=
\frac{1}{2}({\bf 1}-{\bar P}_L\gamma_5+\gamma_5 {\bar P}_T
\slashed t^+)\slashed p_2~,
\ee
where $s_{e^-}$ ($s_{e^+}$) is the polarization vector
and $t^-$ ($t^+$) the transverse beam
polarization 4-vector of $e^-$ ($e^+$), respectively.  
In Eq.~\rf{eq:proje}  (Eq.~\rf{eq:projp})
$P_L$ ($\bar P_L$) [$-1\leq P_L,\bar P_L \leq 1$] denotes the degree
of the longitudinal polarization of $e^-$ ($e^+$) and 
$P_T$ ($\bar P_T$) [$0\leq P_T,\bar P_T \leq 1$] 
denotes the degree of transverse polarization
of $e^-$ ($e^+$), statisfying $P_L^2+P_T^2\leq 1$ and 
$\bar P_L^2+\bar P_T^2\leq 1$.
The $P$ terms are given by 
\be{eq:P}
P=P(\gamma \gamma)+ P(\gamma \tilde{\nu})+P(\gamma Z)+P(Z Z)+
P(Z\tilde{\nu})+P(\tilde{\nu}\tilde{\nu})~,
\ee
where in the following we only list the part which
involves the transverse beam polarization
(the terms not dependent on the beam polarization or the terms which 
depend on the longitudinal beam polarization can be 
found in \cite{Moortgat-Pick:1998sk}):
\be{eq:Pgg}
P(\gamma \gamma)=P_T {\bar P}_T~2 \delta_{ij} e^4
|\Delta(\gamma)|^2 (-r_1)~,
\ee
\be{eq:PgS}
P(\gamma \tilde{\nu})=-P_T {\bar P}_T~\frac{1}{2}\delta_{ij}
e^2 g^2 \Delta(\gamma)\Delta(\tilde{\nu})^*
{\rm Re}\{
V^*_{i1}V_{j1}(r_1-r_2)
\}~,
\ee
\baq{eq:PgZ}
P(\gamma Z)&=& P_T {\bar P}_T~
e^2\delta_{ij}\frac{g^2}{\cos\Theta^2_W}
\Delta(\gamma) \Delta(Z)^*
\nonumber\\[3mm]
&&{}\times
{\rm Re}\{(O'^R_{ij}-O'^L_{ij})
[(L_e+R_e)(-r_1)+(L_e-R_e)r_2 ]\}~,
\eaq
\be{eq:PZZ}
P(Z Z)=P_T {\bar P}_T~\frac{g^4}{\cos\Theta^4_W}|\Delta(Z)|^2
L_e R_e(|O'^L_{ij}|^2+|O'^R_{ij}|^2) (-r_1)~,
\ee
\be{eq:PZS}
P(Z\tilde{\nu})=-P_T {\bar P}_T~\frac{1}{2}\frac{g^4}{\cos\Theta^2_W}
\Delta(Z)\Delta(\tilde{\nu})^* R_e
~{\rm Re}\{V^*_{i1}V_{j1} O'^L_{ij}(r_1-r_2)\}~,
\ee
\be{eq:PSS}
P(\tilde{\nu}\tilde{\nu})=0~,
\ee 
where we have introduced the shorthand notation
\baq{eq:r1}
r_1 &=&[(t^- p_4)(t^+ p_3)+(t^- p_3)(t^+ p_4)]
(p_1 p_2)\nonumber\\[3mm]
&
&+[(p_1 p_4)(p_2 p_3)+(p_1 p_3)(p_2 p_4)-
(p_1 p_2)(p_3 p_4)](t^- t^+)~,
\eaq
\baq{eq:r2}
r_2&=& i~\varepsilon^{\mu\nu\rho\sigma}[
t^+_\mu p_{1\nu} p_{2\rho} p_{4\sigma}
(t^- p_3)+
t^-_\mu p_{1\nu}p_{2\rho} p_{3\sigma}(t^+ p_4)\nonumber\\[3mm]
&& +
t^-_\mu t^+_\nu p_{2\rho} p_{4\sigma}(p_1 p_3)+
t^-_\mu t^+_\nu p_{1\rho}p_{3\sigma} (p_2 p_4)]~,
\eaq
where 
$\Delta(Z)=
i/(s-m^2_Z), 
\Delta(\tilde{\nu})=
i/(t-m^2_{\tilde{\nu}})$, with $s=(p_1+p_2)^2$, 
$t=(p_1-p_4)^2$, $m_{\tilde{\nu}}$ 
($m_Z$)
is the mass of the sneutrino ($Z$ boson) and $\varepsilon^{0123}=1$. 
For the evaluation of the traces we have used 
the FeynCalc package \cite{Mertig:an}.
Note that only terms bilinearly dependent on the
transverse beam polarizations appear for $m_e\to 0$,
since the couplings to $e^+ e^-$ 
are vector- or axialvector-like \cite{Dass:1975mj,Renard,Osland}
(for the $\ti\nu$ exchange the coupling to $e^+ e^-$
can be brought to that form via Fierz identities). 
The cross section for the process \rf{eq:prodchar} is given by
\be{eq:crossectionprod}
{\rm d}\sigma=\frac{1}{2 (2 \pi)^2}
\frac{q}{s^{3/2}}~P~ {\rm d}\cos\theta~ {\rm d}\phi~,
\ee
where $P$ contains the terms for arbitrary beam polarization and
$q$ is the momentum of the $\ti\chi^{\pm}$'s.

Now we consider the $\Sigma^{a,b}$ terms, which means that
we take into account the polarization (or equivalently
the decay) of one of the two produced charginos.
The $\Sigma^a$ term is given by
\begin{equation}
\Sigma^a=
\Sigma^a(\gamma \gamma)+\Sigma^a(\gamma \tilde{\nu})+\Sigma^a(\gamma Z)+
\Sigma^a(Z Z)+\Sigma^a(Z \tilde{\nu})+\Sigma^a(\tilde{\nu}\tilde{\nu})~,
\end{equation}
where in the following we again list only the part which
involves the transverse beam polarization
(for the terms independent of the beam polarization or the terms which 
depend on the longitudinal beam polarization see \cite{Moortgat-Pick:1998sk}):
\be{eq:Sgg}
\Sigma^a(\gamma \gamma)=0~,
\ee
\be{eq:SgS}
\Sigma^a(\gamma \tilde{\nu})=-P_T {\bar P}_T~\frac{1}{2}\delta_{ij}
e^2 g^2 \Delta(\gamma)\Delta(\tilde{\nu})^*
{\rm Re}\{
V^*_{i1}V_{j1}(r^a_1+r^a_2)
\}~,
\ee
\baq{eq:SgZ}
\Sigma^a(\gamma Z)&=& P_T {\bar P}_T~
e^2\delta_{ij}\frac{g^2}{\cos\Theta^2_W}
\Delta(\gamma) \Delta(Z)^*
\nonumber\\[3mm]
&&{}\times
{\rm Re}\{(O'^R_{ij}-O'^L_{ij})
[(L_e+R_e)r^a_1+(L_e-R_e)r^a_2]\}~,
\eaq
\begin{equation}
\Sigma^a(Z Z)=P_T {\bar P}_T~\frac{g^4}{\cos^4\Theta_W}
|\Delta(Z)|^2 L_e R_e (|O'^R_{ij}|^2-|O'^L_{ij}|^2)r^a_1~,
\end{equation}
\be{eq:SZS}
\Sigma^a(Z \tilde{\nu})=-P_T {\bar P}_T~\frac{g^4}{2\cos^2\Theta_W}
\Delta(Z)\Delta(\tilde{\nu})^*
R_e {\rm Re}\{V^*_{i1}V_{j1} O'^L_{ij}(r^a_1+r^a_2)\}~,
\ee
\be{eq:SSS}
\Sigma^a(\tilde{\nu}\tilde{\nu})=0~,
\ee
with
\baq{eq:ra1}
r^a_1&=&-m_{\chi_i}\{[(t^+ p_4)(s^a t^-)+(s^a t^+)
(t^- p_4)](p_1 p_2)\nonumber \\[3mm]
& & +[(s^a p_2)(p_1 p_4)+(s^a p_1)(p_2 p_4)
-(s^a p_4)(p_1 p_2)](t^- t^+)\}
\eaq
\begin{eqnarray}
r^a_2 &=& i~\varepsilon^{\mu\nu\rho\sigma}~ m_{\chi_i}[
t^+_\mu p_{1\nu} p_{2\rho}p_{4\sigma}(s^a t^-)+
t^-_\mu t^+_\nu p_{2\rho}p_{4\sigma}(s^a p_1)
\nonumber\\[3mm]
 & &-
s^a_\mu t^-_\nu p_{1\rho}p_{2\sigma}(t^+ p_4)-
s^a_\mu t^-_\nu t^+_\rho p_{1\sigma}(p_2 p_4)]~,
\end{eqnarray}
where the polarization basis 4-vectors $s^a$
($a=1,2,3$) for $\ti\chi^+_i$
fulfill the orthogonality relations $s^a\cdot s^c=-\delta^{ac}$ 
and $s^a\cdot p_3=0$.
$\Sigma^b$ is obtained by making the replacements 
$s^a\to -s^b, m_{\chi_i} \to m_{\chi_j}, p_4\to p_3$
in Eqs.~\rf{eq:r1} and \rf{eq:ra1}.

The spin density matrices for the decays 
$\ti\chi^-_j\to\ti\nu_{\ell}\ell^-$ 
\rf{eq:lepdec} and $\ti\chi^-_j\to\ti\chi^0_1 W^-$ \rf{eq:Wdec} 
can be written as
\be{eq:decdensity}
(\rho_D)_{\lambda'_j\lambda_j}=D \delta_{\lambda'_j\lambda_j}+
\Sigma^a_D \sigma^a_{\lambda'_j\lambda_j}~,
\ee
where the expansion coefficient are 
\be{eq:Dlep}
D(\ti\nu~\ell)=\frac{g^2}{2} |V_{j1}|^2 (m^2_{\chi_j}-m^2_{\ti\nu})~,
\ee
\be{eq:Slep}
\Sigma^b_D(\ti\nu~\ell)=g^2 |V_{j1}|^2 m_{\chi_j} (s^b \cdot p_{\ell})~,
\ee
for the decay \rf{eq:lepdec}
and 
\baq{eq:DW}
D(\ti\chi^0_1 ~W) &=& g^2 (|O^L_{1j}|^2+|O^R_{1j}|^2)
\left[ \frac{(m^2_{\chi^0_1}+m^2_{\chi_j})m^2_W+
(m^2_{\chi^0_1}-m^2_{\chi_j})^2-2 m^4_W}{2 m^2_W}
\right] \nonumber \\[3mm]
& & - 6 g^2 {\rm Re}({O^L_{1j}}^*O^R_{1j})m_{\chi^0_1} m_{\chi_j}~,
\eaq
\be{eq:SW}
\Sigma^b_D(\ti\chi^0_1 ~W)=g^2  (|O^L_{1j}|^2-|O^R_{1j}|^2) m_{\chi_j}
\left[ \frac{m^2_{\chi_j}-m^2_{\chi^0_1}-2 m^2_W}{m^2_W}
\right]~(s^b \cdot p_W)~,
\ee
for the decay \rf{eq:Wdec}.
Using Eq.~\rf{eq:spindensity} and summing over the
polarization of $\ti\chi^+_i$, whose decay is not
considered, finally gives the cross section for 
$e^+e^-\rightarrow\tilde{\chi}^+_i\tilde{\chi}^-_j\to
\tilde{\chi}^+_i\ell^-\ti\nu_{\ell}~$ $(\tilde{\chi}^+_i W^-\ti\chi^0_1)$:
\be{eq:crossection}
{\rm d}\sigma=\frac{2}{s}\left[P D+\Sigma^a_P \Sigma^a_D\right] 
|\Delta(\ti\chi^-_j)|^2{\rm dLips}~,
\ee
where $P$ and $\Sigma^a_P$ involve the terms for
arbitrary beam polarization.
The Lorentz invariant phase space element ${\rm dLips}$ is given 
in Appendix \ref{appendixB} for the two decays 
\rf{eq:lepdec} and \rf{eq:Wdec}.

\section{Triple product correlations with 
transverse beam polarization\label{sec:3}}

In the following we argue why T-odd observables
as in \rf{eq:Todd} based on triple product correlations
of the sort as in \rf{eq:triple1} and in \rf{eq:triple2}
are expected to vanish at tree-level if at
least one subsequent chargino decay is not observed.

We discuss first the T-odd observables $\langle O^{1,2}_T \rangle$ 
based on the triple product correlations in Eq.~\rf{eq:triple1},
which involve only the production cross section 
$\sigma(e^+e^-\rightarrow\tilde{\chi}^+_i\tilde{\chi}^-_j)$ 
and the quantity $P$, Eq.~\rf{eq:P}.
First we note that these triple products are contained only in 
the kinematic quantity $r_2$ in Eq.~\rf{eq:r2}.
We note further that the two produced charginos should be different
mass eigenstates, i.e. $i\neq j$, otherwise the 
prefactor $V^*_{i1}V_{j1} O'^L_{ij}$ in Eq.~\rf{eq:PZS} would be real.
Then the $\gamma$ exchange does not contribute.
However, it can be shown that also for $i\neq j$ the prefactor 
$V^*_{i1}V_{j1} O'^L_{ij}$ in Eq.~\rf{eq:PZS} is real. 
In fact, this can be verified by a short calculation using
the parametrization for $V$ \cite{Choi:1998ei,Kizukuri:nj}
\begin{equation}
V=
\left(\begin{array}{ccc}
\cos\theta_2 & e^{-i\phi_2}\sin\theta_2\\[4mm]
- e^{i\phi_2}\sin\theta_2 & \cos\theta_2
\end{array}\right)~.
\label{eq:Vchar}
\end{equation}
This result can also be deduced from the formulae given in 
\cite{Choi:2000ta}.
The reason behind lies in the CP property of the quantity $r_2$.
We adopt the method of \cite{Dass:1975mj} to examine the 
behavior of $r_2$ under a CP transfomation.
We first choose the transverse polarizations of 
$e^-$ and $e^+$ either parallel
or anti-parallel to each other (${\bf t}^-={\bf t}^+$ 
or ${\bf t}^-=-{\bf t}^+$). Then the last
two terms in Eq.~\rf{eq:r2} are identical zero.
Applying a CP transformation (in the c.m. system) 
to the first two terms of $r_2$ as follows,
\be{eq:CPtrafo}
  {\bf t}^+({\bf p}_1\times{\bf p}_4)({\bf t}^- \cdot{\bf p}_3)
\stackrel{C}{\longrightarrow}
 {\bf t}^-({\bf p}_2\times{\bf p}_4)({\bf t}^+ \cdot{\bf p}_3)
\stackrel{P}{\longrightarrow}
-  {\bf t}^-({\bf p}_2\times{\bf p}_4)({\bf t}^+\cdot{\bf p}_3)
\ee
one finds that $r_2$ is CP even
(here we sum over the charges of the final charginos so that
${\bf p}_3 \stackrel{C}{\longrightarrow}{\bf p}_3$ and 
${\bf p}_4 \stackrel{C}{\longrightarrow}{\bf p}_4$).
On the other hand $r_2$
is T odd, where T stands for the so-called naive time
reversal (i.e. all momentum and polarization vectors are reversed without
interchanging initial and final state).
Therefore, the prefactor of $r_2$ in Eq.~\rf{eq:r2} vanishes
as a consequence of CPT.
The same conclusion can be derived for the case that the
transverse polarizations of $e^-$ and $e^+$ are orthogonal
to each other. Non-zero contributions to the T-odd observables 
$\langle O^{1,2}_T\rangle$ may arise if terms of the order ${\mathcal O}(m_e)$
are included.
 
As the next step we discuss the triple product correlations
$O^{3,4}_T$, which means we take into account also
the subsequent decay of one of the charginos.
In this case we have to include the terms $\Sigma^a$, 
Eqs.~\rf{eq:Sgg}-\rf{eq:SSS}, which depend on the polarization
of the decaying chargino.
As can be seen, the quantity $r^a_2$ of $\Sigma^a(Z\ti\nu)$
in Eq.~\rf{eq:SZS} is the only term which contains
the triple product correlations \rf{eq:triple2}.
However, its prefactor is again ${\rm Im}(V^*_{i1}V_{j1} O'^L_{ij})$,
which is zero as shown above. 
Unlike in the previous case the reason for the
vanishing prefactor is not a direct consequence of CPT.
In fact, applying a CP transformation to $r^a_2$ in the same manner  
as in the previous case shows that $r^a_2$ is CP odd.
It is also T odd.

Also in the present case there may be non-zero contributions to the T-odd
observables in \rf{eq:Todd} proportional to $m_e$. 
In general non-zero contributions to the T-odd observables
based on the triple products in Eqs.~\rf{eq:triple1} and
\rf{eq:triple2} may also arise by the inclusion of one-loop
contributions. 

Thus we have to conclude that only the $\Sigma^{ab}$ terms contain
non-vanishing triple product correlations with transverse beam polarization.
In order to measure observables based on such
triple product correlations the decays of both 
$\ti\chi^{\pm}$'s must be taken into account \cite{holger}. 
However, in this case transversely polarized beams are
not really necessary, because the same
combinations of CP violating couplings appear in $\Sigma^{ab}$
already in the case of unpolarized beams \cite{karl}.

Although we have derived our results within the MSSM,
we would like to point out that our conclusions remain
valid if SUSY flavour violating terms are included.
In such a case the Lagrange density in Eq.~\rf{eq:lagsneuchar} 
is modified, however, possible CP violating phases from the 
flavour violating sector drop out in the amplitude 
for the $\ti\nu$ exchange (in this context see also \cite{Ghodbane:2000cx}).
This is interesting since in this case $\varphi_{\mu}$
may not be restricted due to the electron electric 
dipole moment (EDM) \cite{Bartl:2003ju}.

\section{Azimuthal asymmetry \label{sec:4}}

\begin{figure}[t]
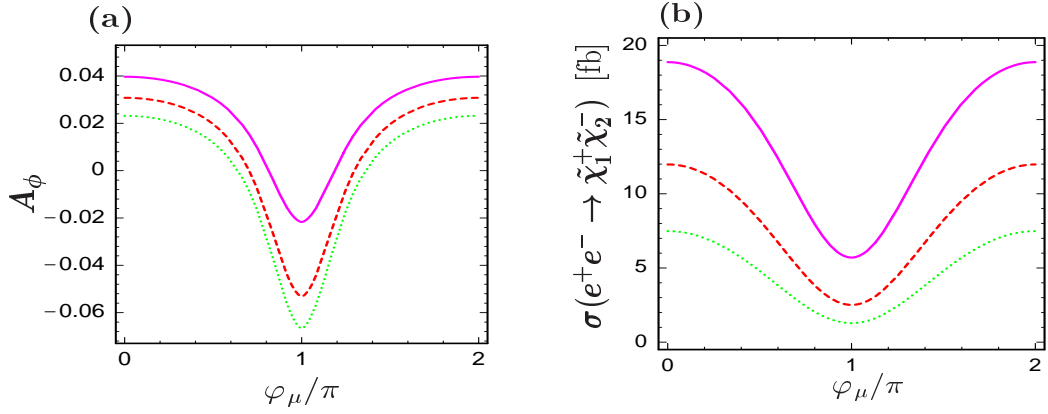

\setlength{\unitlength}{1mm}
\begin{center}
\begin{picture}(150,60)
\put(-50,-145){\mbox{\epsfig{figure=fig.2a,height=22.cm,width=19.4cm}}}
\put(30,-137){\mbox{\epsfig{figure=fig.2b,height=21.cm,width=18.4cm}}}
\end{picture}
\end{center}
\caption{The azimuthal asymmetry in Eq.~\rf{eq:AT1} and the 
cross section $\sigma(e^+ e^-\to \ti\chi^+_1 \ti\chi^-_2)$
as a function of $\varphi_{\mu}$. The three lines correspond to
values of $|\mu|$ (from the top to the bottom) of
$|\mu|=(300,350,400)$~GeV.
The other parameters are chosen as $M_2= 200$~GeV, $\tan\beta = 3$,
$\sqrt{s}=800$~GeV and $m_{\ti\nu}=400$~GeV.}
\label{fig:fig1}
\end{figure}

As we have seen transverse beam polarization does not lead to
a T-odd (CP-odd) observable if chargino production and the decay
of only one of the charginos 
is considered. In order to measure the CP violating parameter 
$\varphi_{\mu}$ and the phase of the $U(1)$ gaugino mass
parameter $\varphi_{M_1}$
in the reaction $e^+e^-\rightarrow\tilde{\chi}^+_i\tilde{\chi}^-_j$
with transverse beam polarization we propose as observable
an azimuthal asymmetry analogous to that of \cite{Choi:2000ta,Chiappetta:1985nb} 
\be{eq:AT}
A_{\phi}=
\frac{\int^+ ({\rm d}\sigma/{\rm d}\phi) {\rm d}\phi- 
\int^- ({\rm d}\sigma/{\rm d}\phi) {\rm d}\phi}{\sigma}~,
\ee
where $\phi$ is the azimuthal angle of the $\tilde{\chi}^{\pm}$'s.
In Eq.~\rf{eq:AT} $\int^{\pm}$ corresponds to an integration over regions where
$\cos 2\phi$ (or $\sin 2\phi$) is positive or negative.
The integration in the numerator has the effect of projecting out 
the terms $\propto P_T \bar P_T$ in the 
formulae for the differential cross section $d\sigma/d\phi$.
Choosing the beam direction along the $z$-axis and 
the transverse polarization of $e^-$ along the $x$-axis
(see Appendix \ref{appendixA}), 
the kinematical factor in Eq.~\rf{eq:r1} can be rewritten as
\be{eq:rer1}
r_1=-\frac{1}{2} q^2 s~ \sin^2\theta
(\sin2\phi \sin\bar{\alpha}+\cos2\phi \cos\bar{\alpha})~,
\ee
where $\bar{\alpha}$ is the angle between 
the transverse polarization vectors of $e^-$ and $e^+$,
and the other quantities are defined in the Appendix A.
This means for the azimuthal asymmetry
that we have two possible integrations depending 
on how the two transverse beam polarizations are
orientated to each other.  
For $\bar{\alpha}=\pi/2$ Eq.~\rf{eq:AT} leads to
\be{eq:AT1}
A_{\phi}=\frac{1}{\sigma}
\left[\int^{\pi/2}_{0} -\int^{\pi}_{\pi/2} 
+\int^{3\pi/2}_{\pi}- \int^{2\pi}_{3\pi/2}
\right]
\frac{{\rm d}\sigma}{{\rm d}\phi} {\rm d}\phi~.
\ee
If we had chosen $\bar{\alpha}=0,\pi$ instead,
the integration over $\phi$ would be in steps of $\pi/4$. 
Under favourable conditions the momentum of the $\ti\chi^{\pm}$'s
can be reconstructed. For such a case we calculate
$A_{\phi}$ and $\sigma$ for $e^+e^-\rightarrow\tilde{\chi}^+_1\tilde{\chi}^-_2$
as a function of $\varphi_{\mu} \in [0,2\pi]$ and the choices
of $|\mu|=(300,350,400)$~GeV, fixing the other parameters as 
$M_2= 200$~GeV, $\tan\beta = 3$, $m_{\ti\nu}=400$~GeV for
$\sqrt{s}=800$~GeV.
We assume that the same degree of transverse polarization 
is feasible as for the longitudinal polarization,
this means we take $P_T=80$\% and $\bar{P}_T=60$\%.  
Fig.~\ref{fig:fig1} shows the result.
As can be seen $A_{\phi}$ depends quite strongly on $\varphi_{\mu}$. We have found
that this dependence gets much weaker for increasing $\tan\beta$, since
in the limit $\tan\beta\to\infty$ the mixing angles and mass eigenvalues
in the chargino sector are independent of $\varphi_{\mu}$. 
We have compared the phase dependence of the  cross section
with the numerical results of \cite{Ghodbane:2000cx} and found agreement.

We have also studied the reaction 
$e^+e^-\rightarrow\tilde{\chi}^+_1\tilde{\chi}^-_1$, and we have 
found the $\varphi_{\mu}$ dependence is much weaker. 
The reason is that the $\varphi_{\mu}$ dependence of the 
denominator and numerator in Eq.~\rf{eq:AT1} almost cancel each other
in a large part of the MSSM parameter space.

The reconstruction of the direction of the $\ti\chi^{\pm}$'s
is not necessary if we consider the subsequent
decays $\ti\chi^-_j\to \ti\nu_{\ell} \ell^- $, Eq.~\rf{eq:lepdec}, 
or $\ti\chi^-_j\to W^-\ti\chi^0_1$, Eq.~\rf{eq:Wdec}, and the 
corresponding azimuthal distribution of $\ell^-$ or $W^-$. 
We define the azimuthal asymmetry according to Eq.~\rf{eq:AT1}
with the cross section given in Eq.~\rf{eq:crossection}. 
Note that only the terms 
Eqs.~\rf{eq:Pgg}-\rf{eq:PSS} and Eqs.~\rf{eq:Sgg}-\rf{eq:SSS}
together with the phase space elements (which are defined in the Appendices)
depend on the azimuthal angle of $\ell^-$ or $W^-$. 

\begin{figure}[t]
\setlength{\unitlength}{1mm}
\begin{center}
\begin{picture}(150,40)
\put(-50,-180){\mbox{\epsfig{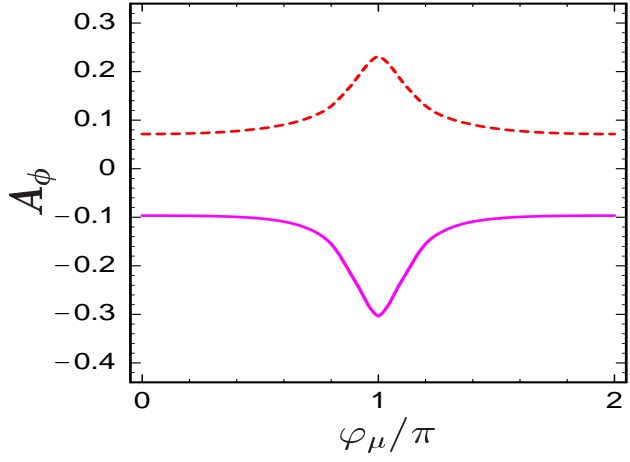}}}
\end{picture}
\end{center}
\caption{The azimuthal asymmetry $A_{\phi}$, Eq.~\rf{eq:AT1}, 
for the reaction $\sigma(e^+ e^-\to \ti\chi^+_1 \ti\chi^-_2)$
at $\sqrt{s}=800$~GeV, with subsequent decays 
$\ti\chi^-_2\to\ti\nu_{\ell}\ell^-$ (solid line) and
$\ti\chi^-_2\to W^-\ti\chi^0_1$ (dashed line)
as a function of $\varphi_{\mu}$. The other parameters are
$|\mu|=400$~GeV, $M_2= 200$~GeV, $\tan\beta = 3$,
$\varphi_{M_1}=0$, $m_{\ti\nu}=150$~GeV.}
\label{fig:fig2}
\end{figure}

In the following calculations of $A_{\phi}$ we assume 
$P_T=80$\% and $\bar{P}_T=60$\%.
In Fig.~\ref{fig:fig2} we show the azimuthal asymmetry, Eq.~\rf{eq:AT1},
of $\ell^-$ and $W^-$ as a function of $\varphi_{\mu} \in [0,2\pi]$.
The MSSM parameters are chosen to be   
$|\mu|=400$~GeV, $M_2= 200$~GeV, $\tan\beta = 3$,
$\varphi_{M_1}=0$, $m_{\ti\nu}=150$~GeV and we will assume
the GUT relation $|M_1|= (5/3) \tan^2\theta_W~ M_2$ throughout.
We vary $\varphi_{\mu}$ over the whole range altough it may 
in general be restricted due to the EDM measurements.
As can be seen in Fig.~\ref{fig:fig2} the CP conserving values
$\varphi_{\mu}=0$ and $\varphi_{\mu}=\pi$ give quite different
results for $A_{\phi}$.
The corresponding cross section is
$\sigma(e^+e^-\rightarrow\tilde{\chi}^+_1\tilde{\chi}^-_2)\approx 29$ fb
for $\varphi_{\mu}=0$ and decreases monotonically for increasing
$\varphi_{\mu}$ until
$\sigma(e^+e^-\rightarrow\tilde{\chi}^+_1\tilde{\chi}^-_2)\approx 1.7$ fb
for $\varphi_{\mu}=\pi$.

\begin{figure}[t]
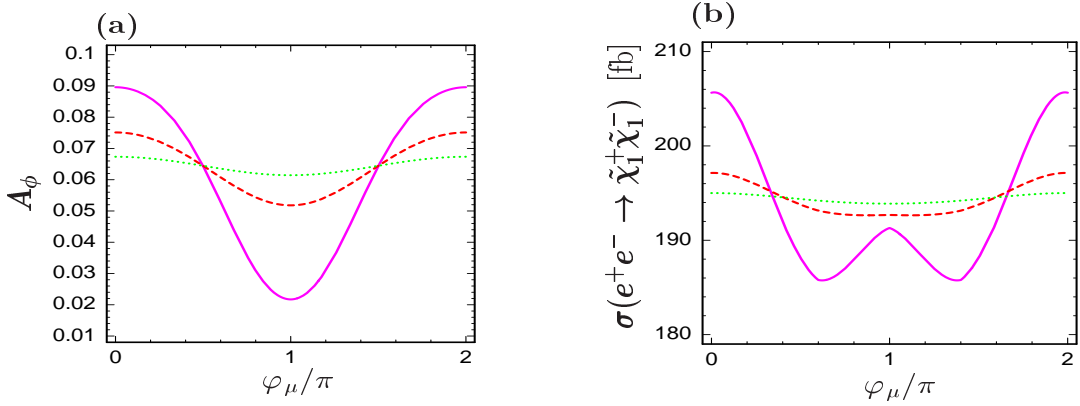

\setlength{\unitlength}{1mm}
\begin{center}
\begin{picture}(150,42)
\put(-50,-145){\mbox{\epsfig{figure=fig.4a,height=21.cm,width=18.4cm}}}
\put(30,-145){\mbox{\epsfig{figure=fig.4b,height=21.cm,width=18.4cm}}}
\end{picture}
\end{center}
\caption{(a) The azimuthal asymmetry $A_{\phi}$, Eq.~\rf{eq:AT1}, for 
$e^+ e^-\to \ti\chi^+_1 \ti\chi^-_1$, 
$\ti\chi^-_1\to \ti\nu_{\ell} \ell^-$
and (b) the cross section for $\sigma(e^+ e^-\to \ti\chi^+_1 \ti\chi^-_1)$
as a function of $\varphi_{\mu}$. 
The three lines correspond to
$\tan\beta=3 ~({\rm solid~line}), 10 ~({\rm dashed~line}), 40~ 
({\rm dotted~line})$,
with $|\mu|=300$~GeV, $M_2= 200$~GeV
and $m_{\ti\nu}=150$~GeV. The c.m. energy is taken to be $\sqrt{s}=500$~GeV.} 
\label{fig:fig3}
\end{figure}

In Fig.~\ref{fig:fig3} we display $A_{\phi}$ and the cross section
for the reaction $e^+e^-\rightarrow\tilde{\chi}^+_1\tilde{\chi}^-_1$,
$\tilde{\chi}^-_1\to\ti\nu_{\ell} \ell^-$, as a function of
$\varphi_{\mu}$ for three values of $\tan\beta=(3,10,40)$,
taking $|\mu|=300$~GeV, $M_2= 200$~GeV and $m_{\ti\nu}=150$~GeV. 
As can be seen the variations of $A_{\phi}$ (Fig.~\ref{fig:fig3}a)
and $\sigma(e^+e^-\rightarrow\tilde{\chi}^+_1\tilde{\chi}^-_1)$ 
(Fig.~\ref{fig:fig3}b)
decrease with increasing $\tan\beta$.
For low $\tan\beta (=3)$ again the CP conserving points lead to quite
different results for $A_{\phi}$.
Also the cross section depends in a significant way on $\varphi_{\mu}$,
for example, for $\tan\beta=3$ the absolut minimum of the cross section is
reached for CP violating points $\varphi_{\mu}\approx 
\frac{3}{5}\pi,\frac{7}{5}\pi$ 
(see Fig.~\ref{fig:fig3}b).

\begin{figure}[t]
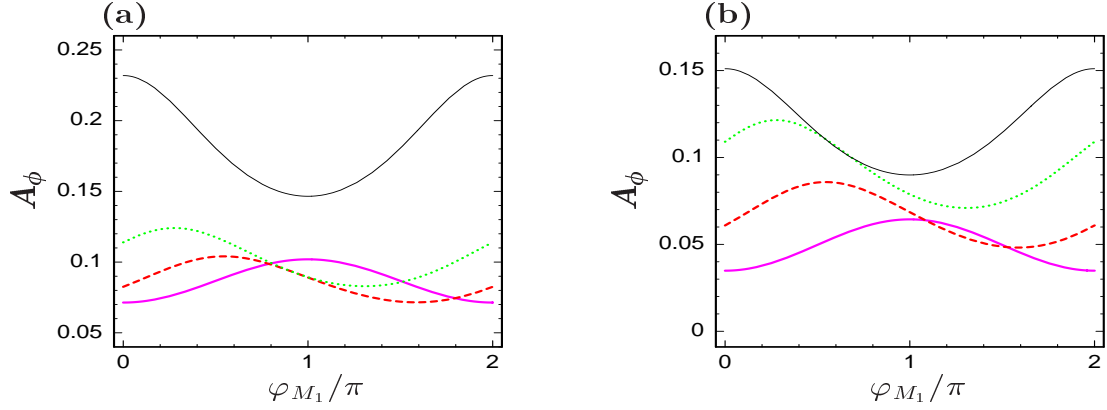

\setlength{\unitlength}{1mm}
\begin{center}
\begin{picture}(150,35)
\put(-50,-145){\mbox{\epsfig{figure=fig.5a,height=22.cm,width=19.4cm}}}
\put(30,-145){\mbox{\epsfig{figure=fig.5b,height=22.cm,width=19.4cm}}}
\end{picture}
\end{center}
\caption{The azimuthal asymmetry $A_{\phi}$, Eq.~\rf{eq:AT1}, for
the reaction $e^+e^-\rightarrow\tilde{\chi}^+_1\tilde{\chi}^-_2$,
$\tilde{\chi}^-_2\to\ti\chi^0_1 W^-$ as a function of $\varphi_{M_1}$ for
$\sqrt{s}=800$~GeV, $|\mu|=400$~GeV, $M_2= 200$~GeV, $\tan\beta = 3$,
$\varphi_{\mu}=
0 ~({\rm thick~solid~line}),\pi/2~({\rm dashed~line}),3\pi/4
~({\rm dotted~line}),\pi~({\rm thin~solid~line})$.
(a) shows $A_{\phi}$ for $m_{\ti\nu}=150$~GeV
and (b) shows $A_{\phi}$ for $m_{\ti\nu}=2$~TeV.}
\label{fig:fig4}
\end{figure}

In Fig.~\ref{fig:fig4} we plot $A_{\phi}$ 
for the reaction $e^+e^-\rightarrow\tilde{\chi}^+_1\tilde{\chi}^-_2$,
with $\tilde{\chi}^-_2\to W^-\ti\chi^0_1$.
We fix $|\mu|=400$~GeV, $M_2= 200$~GeV, $\tan\beta = 3$
and vary $\varphi_{M_1}\in [0,2\pi]$ for $\varphi_{\mu}=
(0,\pi/2,3\pi/4,\pi)$.
Fig.~\ref{fig:fig4}a shows $A_{\phi}$ for $m_{\ti\nu}=150$~GeV
and Fig.~\ref{fig:fig4}b for $m_{\ti\nu}=2$~TeV.
For $m_{\ti\nu}=2$~TeV (and assuming that the other sfermion masses 
of the first two generations are also heavy) the restriction
from the EDMs on $\varphi_{\mu}$ is relaxed.
One sees that $A_{\phi}$ depends quite strongly on the 
CP violating phases $\varphi_{M_1}$ and $\varphi_{\mu}$.
Note that the $\varphi_{M_1}$ dependence is due to the
decay amplitude.

\begin{figure}[t]
\setlength{\unitlength}{1mm}
\begin{center}
\begin{picture}(150,60)
\put(-50,-180){\mbox{\epsfig{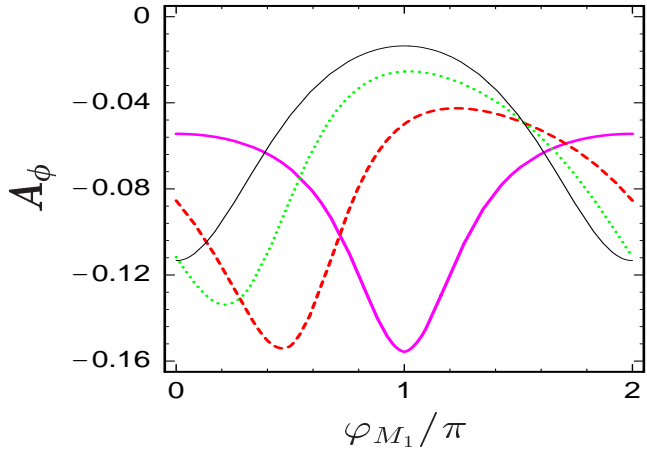}}}
\end{picture}
\end{center}
\caption{The azimuthal asymmetry $A_{\phi}$, Eq.~\rf{eq:AT1}, for
the reaction $e^+e^-\rightarrow\tilde{\chi}^+_1\tilde{\chi}^-_1$ 
($\sqrt{s}=800$~GeV) with subsequent decay $\tilde{\chi}^-_1\to\ti\chi^0_1 W^-$
as a function of $\varphi_{M_1}$.
We fix $|\mu|=400$~GeV, $M_2= 400$~GeV, $\tan\beta = 3$, $m_{\ti\nu}=150$~GeV,
$\varphi_{\mu}=
0 ~({\rm thick~solid~line}),\pi/2~({\rm dashed~line}),3\pi/4
~({\rm dotted~line}),\pi~({\rm thin~solid~line})$.}
\label{fig:fig5}
\end{figure}

In Fig.~\ref{fig:fig5} we plot $A_{\phi}$ 
for the reaction $e^+e^-\rightarrow\tilde{\chi}^+_1\tilde{\chi}^-_1$
with the subsequent decay $\tilde{\chi}^-_1\to\ti\chi^0_1 W^-$.
The MSSM parameters are chosen to be $|\mu|=400$~GeV, $M_2=400$~GeV, $\tan\beta = 3$.
We vary $\varphi_{M_1}\in [0,2\pi]$ for $\varphi_{\mu}=(0,\pi/2,3\pi/4,\pi)$.
As can be seen also in this case the phase dependence of 
the azimuthal asymmetry of the $W$ boson is very pronounced.

\section{Conclusion \label{sec:5}}

We have considered the process $e^+e^-\to\ti\chi^+_i\ti\chi^-_j$
at a linear collider with transversely polarized $e^+$ and $e^-$ beams.
We have given the analytical expressions for the cross section of
these processes in the spin density matrix formalism.
We have given arguments why triple product correlations involving the
transverse $e^{\pm}$ polarizations vanish if at least
one subsequent chargino decay is not observed.
Our framework has been the MSSM, but this statement
is also valid for the general MSSM with SUSY flavour violation.
We have proposed and studied azimuthal asymmetries 
in the processes $e^+e^-\rightarrow\tilde{\chi}^+_i\tilde{\chi}^-_j\to
\tilde{\chi}^+_i\ell^-\ti\nu_{\ell}~$ $(\tilde{\chi}^+_i W^-\ti\chi^0_1)$.
We have demonstrated that these azimuthal asymmetries are well 
suited to investigate the effect of the SUSY CP phases 
$\varphi_{\mu}$ and $\varphi_{M_1}$.

\section*{Acknowledgements}

We thank P.~Osland and A.~Vereshagin for communicating their
results prior to publication and for important discussions.
Useful discussions with H.~Eberl, H.~Fraas, S.~Hesselbach, 
O.~Kittel, W.~Majerotto and G.~Moortgat-Pick are also gratefully acknowledged.
A.B. is grateful to the organizers of the 8th Workshop on High
Energy Physics Phenomenology at Mumbai, India, for kind hospitality
and to B.~Ananthanarayan, R.~Godbole and S.~Rindani for enlightening discussions.  
This work is supported by the `Fonds zur
F\"orderung der wissenschaftlichen Forschung' (FWF) of Austria, project
No. P16592-N02 and by the European Community's 
Human Potential Programme
under contract HPRN--CT--2000--00149.

\section*{Appendix}

\begin{appendix}

\section{Momentum and polarization vectors\label{appendixA}}

We define the transverse beam polarization 4-vectors
in Eqs.~\rf{eq:proje} and \rf{eq:projp} as
$t^-=\cos\alpha n_1+\sin\alpha n_2$ and 
$t^+=\cos{\bar\alpha} n_1+\sin{\bar\alpha} n_2$.
We choose the $z$-axis along the beam direction in the c.m. system, 
and $n_1=(0,1,0,0)$, $n_2=(0,0,1,0)$.
Without loss of generality, we take $\alpha=0$ throughout. 
The 4-momenta of the $\ti\chi^{\pm}$'s are given by
\be{eq:momentumchar}
p_{\chi_j}=p_4=q
(E_{\chi_j}/q,\cos\phi \sin\theta,\sin\phi \sin\theta,\cos\theta)~,
\ee
with
\be{eq:energy}
E_{\chi_{i,j}}=\frac{s+m^2_{\chi_{i,j}}-m^2_{\chi_{j,i}}}{2 \sqrt{s}}~,\qquad
q=\frac{\lambda^{\frac{1}{2}}(s,m^2_{\chi_i},m^2_{\chi_j})}{2 \sqrt{s}}~,
\ee
where $\lambda(a,b,c)=a^2+b^2+c^2-2(a b + a c + b c)$.
The three spin basis vectors of $\ti\chi^-_j$
are chosen to be 
\baq{eq:polvec}
s^1_{\chi_j}&=&\left(0,\frac{{\bf s}_2\times{\bf s}_3}
{|{\bf s}_2\times{\bf s}_3|}\right)=
(0,\cos\phi \cos\theta,\sin\phi \cos\theta,-\sin\theta)~,
\nonumber \\[3mm]
s^2_{\chi_j}&=&\left(0,
\frac{{\bf p}_{e^-}\times{\bf p}_{\chi_j}}
{|{\bf p}_{e^-}\times{\bf p}_{\chi_j}|}\right)=
(0,-\sin\phi,\cos\phi,0)~,
\nonumber \\[3mm]
s^3_{\chi_j}&=&\frac{1}{m_{\chi_j}}
\left(q, 
\frac{E_{\chi_j}}{q}{\bf p}_{\chi_j} \right)=
\frac{E_{\chi_j}}{m_{\chi_j}}
(q/E_{\chi_j},\cos\phi \sin\theta,\sin\phi \sin\theta,\cos\theta)~.
\eaq
The 4-momentum of the lepton in the decay $\ti\chi^-_j
\to\ti\nu_{\ell} \ell^-$ is given by
\be{eq:fourlep}
p_{\ell}=
|{\bf p_{\ell}}| (1,\cos\phi_1 \sin\theta_1,\sin\phi_1 \sin\theta_1,
\cos\theta_1)~,
\ee
where
\be{eq:lepmom}
|{\bf p_{\ell}}|=\frac{m^2_{\chi_j}-
m^2_{\ti\nu}}{2(E_{\chi_j}-q\cos\vartheta)}~,
\ee
and
\be{eq:angle}
\cos\vartheta=\sin\theta \sin\theta_1 \cos(\phi-\phi_1)+
\cos\theta \cos\theta_1~.
\ee
The 4-momentum of the $W$ in the decay $\ti\chi^-_j
\to\ti\chi^0_1 W^-$ is given by
\be{eq:momentumdecay}
p_W=
(E_W,|{\bf p}_{W}|\cos\phi_1 \sin\theta_1,
|{\bf p}_{W}| \sin\phi_1 \sin\theta_1,|{\bf p}_{W}| \cos\theta_1)~.
\ee
with
\baq{eq:Wmom}
|{\bf p}^{\pm}_W|&=&\left[
2|{\bf p}_{\chi_j}|^2 (1-\cos^2\vartheta)+2 m^2_{\chi_j}
\right]^{-1}
\left[
(m^2_{\chi_j}+m^2_W-m^2_{\chi^0_1})|{\bf p}_{\chi_j}|
\cos\vartheta \right. \nonumber \\[3mm]
&& {}
\left. \pm E_{\chi_j}\sqrt{\lambda(m^2_{\chi_j},m^2_W,m^2_{\chi^0_1})-
4|{\bf p}_{\chi_j}|^2~m^2_W~(1-\cos^2\vartheta)}
\right]~.
\eaq
There are two solutions $|{\bf p}^{\pm}_W|$ if 
$|{\bf p}^0_{\chi_j}|<|{\bf p}_{\chi_j}|$,
where $|{\bf p}^0_{\chi_j}|= 
\lambda^{\frac{1}{2}}(m^2_{\chi_j},m^2_W,m^2_{\chi^0_1})/2 m_W$  
is the chargino momentum if the $W$ boson is produced at rest.
The $W$ decay angle $\vartheta$ is constrained in that case 
and the maximal angle $\vartheta^{\rm max}$ is given by
\be{eq:maxangle}
\sin\vartheta^{\rm max}=
\frac{|{\bf p}^0_{\chi_j}|}{|{\bf p}_{\chi_j}|}=
\frac{\sqrt{s}}{m_W}
\frac{\lambda^{\frac{1}{2}}(m^2_{\chi_j},m^2_W,m^2_{\chi^0_1})}
{\lambda^{\frac{1}{2}}(s,m^2_{\chi_i},m^2_{\chi_j})}\leq 1~.
\ee
If $|{\bf p}^0_{\chi_j}|>|{\bf p}_{\chi_j}|$, the
decay angle $\vartheta$ is not constrained and there is
only the physical solution $|{\bf p}^+_W|$.

\section{Phase space \label{appendixB}}

The Lorentz invariant phase space element in Eq.~\rf{eq:crossection} is given by
\be{eq:phasespace}
{\rm d Lips} =\frac{1}{2 \pi} {\rm d Lips}(s,p_{\chi_i},p_{\chi_j})
{\rm d}s_{\chi_j}
{\rm d Lips}(s_{\chi_j},p_{\ti\nu},p_{\ell})
\ee
for the subsequent decay $\ti\chi^-_j\to\ti\nu_{\ell} \ell^-$ \rf{eq:lepdec}, and by
\be{eq:phasespace1}
{\rm d Lips} =\frac{1}{2 \pi} {\rm d Lips}(s,p_{\chi_i},p_{\chi_j})
{\rm d}s_{\chi_j}
\sum_{\pm} {\rm d Lips}(s_{\chi_j},p_{\chi^0_1},p^{\pm}_{W})
\ee
for the subsequent decay $\ti\chi^-_j\to\ti\chi^0_1 W^-$ \rf{eq:Wdec}. 
The Lorentz invariant phase space elements in Eq.~\rf{eq:phasespace}
and Eq.~\rf{eq:phasespace1} read
\be{eq:prodphs}
{\rm d Lips}(s,p_{\chi_i},p_{\chi_j})=\frac{1}{4(2 \pi)^2} \frac{q}{\sqrt{s}}
\sin\theta~ {\rm d}\theta~ {\rm d}\phi
\ee
\be{eq:lepphs}
{\rm d Lips}(s_{\chi_j},p_{\ti\nu},p_{\ell})=
\frac{1}{2(2 \pi)^2}\frac{|{\bf p}_{\ell}|^2}{m^2_{\chi_j}-m^2_{\ti\nu}}
\sin\theta_1~ {\rm d}\theta_1~ {\rm d}\phi_1~,
\ee
\be{eq:dlips2}
{\rm d Lips}(s_{\chi_j},p_{\chi^0_1},p^{\pm}_W)=\frac{1}{4(2\pi)^2}~
\frac{|{\bf p}^{\pm}_W|^2}{|E_W^{\pm}~|{\bf p}_{\chi_j}|\cos\vartheta-
E_{\chi_j}~|{\bf p}^{\pm}_W||}~
\sin\theta_1~ {\rm d}\theta_1~ {\rm d}\phi_1~.
\ee

\end{appendix}

\end{document}